\documentclass[aps,prd,twocolumn,showpacs,11pt,tightenlines]{revtex4-1}
\usepackage{url}
\usepackage{bm}
\usepackage{amsmath}
\usepackage{graphicx}
\usepackage{SIunits}

\DeclareMathOperator{\br}{Br}

\pdfoutput=1

\preprint{\parbox[t]{0.5\textwidth}{\raggedleft
   TTP14-008, KA-TP-09-2014, SFB/CPP-14-17, IPPP/14/21, DCPT/14/42}}
\allowdisplaybreaks

\begin{document} 

\title{Benchmarks for Higgs Pair Production and Heavy Higgs Searches\\
  in the Two-Higgs-Doublet Model of Type II}

\author{Julien Baglio}
\email{julien.baglio@kit.edu}
\affiliation{Institut f\"{u}r Theoretische Physik,
Karlsruhe Institute of Technology, Engesserstra\ss{}e 7,
D-76128 Karlsruhe, Germany}

\author{Otto Eberhardt}
\email{otto.eberhardt@roma1.infn.it}
\affiliation{Istituto Nazionale di Fisica Nucleare,
Sezione di Roma, Piazzale Aldo Moro 2, I-00185 Roma, Italy}

\author{Ulrich Nierste}
\email{ulrich.nierste@kit.edu}
\affiliation{Institut f\"{u}r Theoretische Teilchenphysik,
Karlsruhe Institute of Technology, Engesserstra\ss{}e 7,
D-76128 Karlsruhe, Germany}

\author{Martin Wiebusch}
\email{martin.wiebusch@durham.ac.uk}
\affiliation{Institute for Particle Physics and Phenomenology,
Durham University, Durham DH1 3LE, United Kingdom}

\date{May 28, 2014}

\begin{abstract}
  The search for additional Higgs particles and the exact measurements
  of Higgs (self-) couplings is a major goal of future collider experiments. In
  this paper we investigate the possible sizes of new physics signals in these
  searches in the context of the $CP$-conserving two-Higgs doublet model of type
  II. Using current constraints from flavour, electroweak precision, and Higgs
  signal strength data, we determine the allowed sizes of the triple Higgs
  couplings and the branching fractions of the heavy Higgs bosons into lighter
  Higgs bosons. Identifying the observed Higgs resonance with the light
  $CP$-even 2HDM Higgs boson $h$, we find that the $hhh$ coupling cannot exceed
  its SM value, but can be reduced by a factor of $0.56$ at the \unit{2}{\sigma}
  level. The branching fractions of the heavy neutral Higgs bosons $H$ and $A$
  into two-fermion or two-vector-boson final states can be reduced by factors of
  $0.4$ and $0.01$, respectively, if decays into lighter Higgs boson are
  possible and if the mass of the decaying Higgs is below the $t\bar t$
  threshold. To facilitate future studies of collider signatures in 2HDM
  scenarios with large triple Higgs couplings or decay modes of the heavy Higgs
  bosons not covered by the SM Higgs searches we provide a set of benchmark
  points which exhibit these features and agree with all current constraints. We
  also discuss the effect of the heavy Higgs bosons on the $gg\to hh$ cross
  section at a \unit{14}{TeV} LHC for some of these benchmarks. For $m_H$ below
  the $hh$ threshold we see a reduction of the SM $gg\to hh$ cross section due
  to destructive interference, but for $m_H$ above the $hh$ threshold current
  constraints allow enhancement factors above 50. An enhancement factor of 6 is
  still possible in scenarios in which the heavy Higgs particles would not be
  discovered by standard searches after \unit{300}{\power{fb}{-1}} of data.
\end{abstract}

\maketitle

\section{Introduction}
\label{sec:intro}

The discovery of a Higgs boson with a mass of \unit{126}{GeV} is the undisputed
highlight of Run I of the Large Hadron Collider (LHC) at CERN
\cite{Chatrchyan:2012ufa, Aad:2012tfa}. The measured signal strengths agree with
the Standard Model (SM) predictions in several production and decay channels of
the discovered particle \cite{ATLAS-CONF-2013-034, Chatrchyan:2013lba},
establishing the Higgs mechanism \cite{Higgs:1964pj, Englert:1964et,
  Guralnik:1964eu} as the correct theoretical framework of elementary particle
mass generation.  This achievement has been recognised with the 2013 Nobel Prize
for Physics for Peter Higgs and Fran\c{c}ois Englert. However, it is not clear
at all whether the Higgs sector is indeed \emph{minimal}, containing only a
single Higgs doublet.  The most straightforward way to go beyond the SM is the
addition of a second Higgs doublet \cite{Lee:1973iz} to the field content of the
model. In this paper we focus on the $CP$-conserving two-Higgs-doublet-model
(2HDM) of type II and identify the observed \unit{126}{GeV} state as the
lightest $CP$-even Higgs boson $h$.  Currently a lot of effort is spent to
confront the above-described model to LHC data \cite{Ferreira:2011aa,
  Blum:2012kn, Basso:2012st, Cheon:2012rh, Carmi:2012in, Ferreira:2012nv,
  Drozd:2012vf, Chang:2012zf, Chen:2013kt, Celis:2013rcs, Giardino:2013bma,
  Grinstein:2013npa, Shu:2013uua, Barroso:2013zxa, Belanger:2013xza,
  Barger:2013ofa, Lopez-Val:2013yba, Bhattacharyya:2013rya, Eberhardt:2013wia,
  Chang:2013ona, Celis:2013ixa, Wang:2013sha}. Remarkably, current data do not
at all push the model parameters to the decoupling limit $m_{A,H,H^\pm}\to
\infty$, but permit $A$ and $H$ masses well below \unit{200}{GeV}
\cite{Eberhardt:2013uba, Celis:2013ixa, Cheung:2013rva, Arhrib:2013oia}

\begin{figure}
  \centering
  \includegraphics{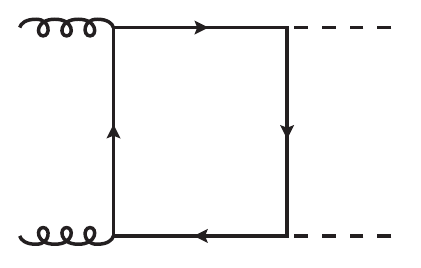} \qquad \qquad
  \includegraphics{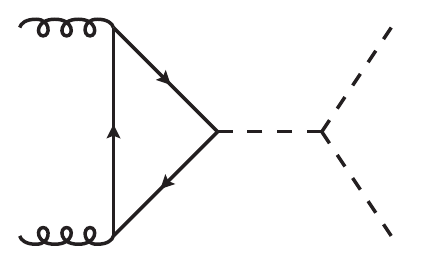}
  \caption{Representative diagrams contributing to $gg\to hh$. Other diagrams
    are obtained by permuting the external lines. In the 2HDM both $h$ and $H$
    can be exchanged between the top loop and the $h$ pair in the right
    diagram.\label{fig:diag}}
\end{figure}

A major target of the future LHC programme is the search for Higgs pair
production. In the gluon fusion channel, which is the dominant production
process, there are two classes of diagrams at leading order contributing to this
process, depicted in Fig.~\ref{fig:diag}.
Higgs pair production probes the triple-$h$ coupling which in the SM is entirely
fixed in terms of the Higgs mass and vacuum expectation value. The corresponding
diagram (Fig.~\ref{fig:diag}b) is smaller than the box diagram by roughly a
factor of 3, depending on the invariant mass of the Higgs pair.

In models with extended Higgs sectors the value of the triple-$h$ coupling can
differ from its SM value and additional Higgs bosons can be exchanged in the
$s$-channel. Higgs pair production at hadron and linear colliders has been
studied extensively in the literature \cite{Djouadi:1999gv, Djouadi:1999rca,
  Baur:2003gp, Dolan:2012rv, Papaefstathiou:2012qe, Baglio:2012np,
  Goertz:2013kp, Barr:2013tda}, and QCD corrections to the SM $gg\to hh$ cross
section, which is the dominant process, are known to next-to-leading order in
the infinite top mass approximation \cite{Dawson:1998py}, to subleading orders
in $1/m_t$ \cite{Grigo:2013rya} and beyond next-to-leading order
\cite{Shao:2013bz, deFlorian:2013uza, deFlorian:2013jea}. Parton-shower effects
have been studied in \cite{Frederix:2014hta}. A recent study
\cite{Barger:2013jfa} has found that within the SM the triple-$h$ coupling can
be measured with an accuracy of $40\%$ at the LHC with an integrated luminosity
of \unit{3}{\power{ab}{-1}} collected at an energy of \unit{14}{TeV}. Other
studies have addressed extended Higgs sectors \cite{Arhrib:2009hc, Dolan:2012ac}
and the sensitivity of the triple Higgs coupling to new physics
\cite{Efrati:2014uta}.

The purpose of this paper is to use all current constraints on the
type-II 2HDM parameter space to determine the allowed ranges of the
triple-Higgs couplings and the branching fractions of the heavy Higgs
bosons in order to prepare the ground for collider studies of the
various Higgs pair production and decay channels. Specifically, we
address the following questions:
\begin{itemize}
\item[i)] To what extent can the triple-$h$ coupling deviate from its
  SM value? 
\item[ii)] What are the maximal sizes of the triple-Higgs couplings
  involving $H$, $A$ or $H^\pm$?
\item[iii)] Can decays like $H\to hh$ or $A\to Zh$ lead to a large suppression
  of the branching fractions in the standard searches for $H$ and $A$?
\item[iv)] To what extent can $gg\to hh$ be enhanced with respect to
  the SM prediction?
\end{itemize}
To facilitate future detailed studies of collider signatures we provide a set of
benchmark points which maximise the above-mentioned features, but still agree
with all experimental and theoretical constraints.

This paper is organised as follows: in Section~\ref{sec:model} we recall the
considered model and fix our notations.  Section~\ref{sec:constr} discusses the
input to our analysis, namely Higgs signal strengths, electroweak precision
observables, and relevant flavour observables.  Section~\ref{sec:results}
addresses questions i) and ii) above.  Questions iii) and iv) are discussed in
Section~\ref{sec:phen}, where we define the benchmark scenarios.
In Section~\ref{sec:conclusions} we present our conclusions.

\section{The Two-Higgs-Doublet Model of Type II}
\label{sec:model}

In this paper we consider the $CP$-conserving two-Higgs-doublet-model
of type II with a softly broken $Z_2$ symmetry. For details about this
model we refer to \cite{Gunion:2002zf}, whose notations we adopt
here. The Higgs potential is given by
\begin{align}
 V
 &=m_{11}^2\Phi_1^\dagger\Phi_1^{\phantom{\dagger}}
   +m_{22}^2\Phi_2^\dagger\Phi_2^{\phantom{\dagger}}
   -m_{12}^2 ( \Phi_1^\dagger\Phi_2^{\phantom{\dagger}}
              +\Phi_2^\dagger\Phi_1^{\phantom{\dagger}})
  \nonumber \\
 &\phantom{{}={}}+\tfrac12 \lambda_1(\Phi_1^\dagger\Phi_1^{\phantom{\dagger}})^2
   +\tfrac12 \lambda_2(\Phi_2^\dagger\Phi_2^{\phantom{\dagger}})^2
 \nonumber \\
 &\phantom{{}={}}
  +\lambda_3(\Phi_1^\dagger\Phi_1^{\phantom{\dagger}})
            (\Phi_2^\dagger\Phi_2^{\phantom{\dagger}})
  +\lambda_4(\Phi_1^\dagger\Phi_2^{\phantom{\dagger}})
            (\Phi_2^\dagger\Phi_1^{\phantom{\dagger}})
 \nonumber \\
 &\phantom{{}={}}
  +\tfrac12 \lambda_5[ (\Phi_1^\dagger\Phi_2^{\phantom{\dagger}})^2
                      +(\Phi_2^\dagger\Phi_1^{\phantom{\dagger}})^2].\label{eq:pot}
\end{align}
The physical scalar spectrum of this model consists of two $CP$-even
neutral scalars $h$ and $H$, a $CP$-odd neutral scalar $A$ and a
charged scalar $H^\pm$. The masses of these states are denoted as
$m_\phi$ with $\phi\in{h,H,A,H^\pm}$. Throughout this paper we assume
that the light $CP$ even scalar $h$ is the observed Higgs resonance
and keep $m_h=\unit{126}{GeV}$ fixed. For the remaining independent
real parameters of the model we choose 
\begin{align}\label{eq:params}
  &\tan\beta=v_2/v_1,
  \;
  \beta-\alpha,
  \;
  m_{12}^2,
 \nonumber \\
  &m_H,
  \;
  m_A,
  \;
  m_{H^\pm},
\end{align}
where $v_1/\sqrt{2}$ and $v_2/\sqrt{2}$ denote the vacuum expectation values of
the neutral components of $\Phi_1$ and $\Phi_2$, respectively, and $\alpha$
denotes the mixing angle of the two $CP$-even neutral Higgs bosons. In this
parametrisation the tree-level couplings of the Higgs bosons to SM vector bosons
and fermions only depend on $\tan\beta$ and $\beta-\alpha$. The couplings of the
light $CP$-even Higgs boson $h$ are SM-like for $\beta-\alpha=\pi/2$. We call
this the \emph{alignment limit}. To express the quartic couplings $\lambda_i$ in
terms of the physical parameters \eqref{eq:params} we use the tree-level
relation
\begin{equation}
    v_1^2+v_2^2 = v^2
  = \frac{M_W^2}{\pi\alpha_\text{em}}
    \left(1-\frac{M_W^2}{M_Z^2}\right)
\end{equation}
with $M_Z=\unit{91.1878}{GeV}$, $M_W=\unit{80.3693}{GeV}$ and
$\alpha_\text{em}\equiv\alpha_\text{em}(M_Z)=1/128.9529$
\cite{Eberhardt:2012gv}.

To compare the different triple Higgs couplings in the 2HDM with the
SM triple Higgs coupling we define the ratios
\begin{equation}
  c_{\phi_1\phi_2\phi_3}=
  \frac{g^\text{2HDM}_{\phi_1\phi_2\phi_3}}{g^\text{SM}_{hhh}}
  \; ,
\end{equation}
where $\phi_1,\phi_2,\phi_3\in\{h,H,A,H^\pm\}$,
$g^\text{2HDM}_{\phi_1\phi_2\phi_3}$ denotes the corresponding 2HDM
triple Higgs coupling and $g^\text{SM}_{hhh}$ denotes the SM triple
Higgs coupling with a fixed SM Higgs mass of \unit{126}{GeV}. In terms
of the parameters in \eqref{eq:params} these ratios are
\begin{widetext}
\begin{align}
 c_{hhh} & = \frac{\cos ^3\alpha }{\sin \beta} 
             -\frac{\sin ^3\alpha }{\cos \beta} 
             -\frac{m_{12}^2}{m_h^2}\cos(\beta-\alpha) 
             \left( \frac{\cos ^2\alpha}{\sin ^2\beta} 
             -\frac{\sin ^2\alpha }{\cos ^2 \beta} \right)
 \; ,\label{eq:triple}\\
 c_{HHH} & = \frac{m_H^2}{m_h^2}
             \left( \frac{\sin ^3\alpha }{\sin \beta} 
             +\frac{\cos ^3\alpha }{\cos \beta}\right) 
             -\frac{m_{12}^2}{m_h^2} \sin (\beta -\alpha )
             \left( \frac{\cos ^2\alpha}{\cos ^2\beta} 
             -\frac{\sin ^2\alpha}{\sin ^2\beta}\right)
 \; ,\nonumber\\
 c_{hhH} & = \frac{\cos(\beta-\alpha)}{\cos \beta \sin \beta}
             \left[ \frac{\cos \alpha \sin \alpha}{3}
             \left( 2+\frac{m_H^2}{m_h^2}\right) 
             +\frac{m_{12}^2}{m_h^2}
             \left(\frac{1}{3}
             -\frac{\cos \alpha \sin \alpha}{\cos \beta \sin \beta}
	     \right) \right]
 \; ,\nonumber\\
 c_{hHH} & = \frac{\sin(\beta-\alpha)}{\cos \beta \sin \beta}
             \left[ -\frac{\cos \alpha \sin \alpha}{3}
             \left( 1+2\frac{m_H^2}{m_h^2}\right) 
             +\frac{m_{12}^2}{m_h^2}\left(\frac{1}{3}
             +\frac{\cos \alpha \sin \alpha}{\cos \beta \sin \beta}
             \right) \right]
 \; ,\nonumber\\
 c_{hH^+H^-} & = \frac{1}{3}\left[ -\sin(\beta-\alpha)
                 \left( 1-2\frac{m_{H^\pm}^2}{m_h^2}\right) 
                 +\left( \frac{\cos \alpha}{\sin \beta}
                 -\frac{\sin \alpha}{\cos \beta}\right) 
                 \left( 1- \frac{1}{\cos \beta \sin \beta}
                 \frac{m_{12}^2}{m_h^2}\right) \right] 
 \; ,\nonumber\\
 c_{HH^+H^-} & = \frac{1}{3}\left[ -\cos(\beta-\alpha)
                 \left( \frac{m_H^2}{m_h^2}
                 -2\frac{m_{H^\pm}^2}{m_h^2}\right) 
                 +\left( \frac{\cos \alpha}{\cos \beta}
                 +\frac{\sin \alpha}{\sin \beta}\right) 
                 \left( \frac{m_H^2}{m_h^2}
                 -\frac{1}{\cos \beta \sin \beta}
                 \frac{m_{12}^2}{m_h^2}\right) \right]
 \; ,\nonumber\\
 c_{hAA} & = \frac{1}{3}\left[ -\sin(\beta-\alpha)
             \left( 1-2\frac{m_A^2}{m_h^2}\right) 
             +\left( \frac{\cos \alpha}{\sin \beta}
             -\frac{\sin \alpha}{\cos \beta}\right) 
             \left( 1- \frac{1}{\cos \beta \sin \beta}
             \frac{m_{12}^2}{m_h^2}\right) \right]
 \; ,\nonumber\\
 c_{HAA} & = \frac{1}{3}\left[ -\cos(\beta-\alpha)
             \left( \frac{m_H^2}{m_h^2}-2\frac{m_A^2}{m_h^2}\right)
             +\left( \frac{\cos \alpha}{\cos \beta}
             +\frac{\sin \alpha}{\sin \beta}\right)
             \left( \frac{m_H^2}{m_h^2}
             -\frac{1}{\cos \beta \sin \beta}
             \frac{m_{12}^2}{m_h^2}\right) \right]
 \; .\nonumber
\end{align}
\end{widetext}
These formluae agree with the triple Higgs couplings from \cite{Boudjema:2001ii}.
In the alignment limit $\beta-\alpha=\frac\pi2$ these expressions
simplify to
\begin{align}
 c_{hhh} & = 1,\nonumber\\
 c_{hhH} & = 0,\nonumber\\
 c_{hXX} & = \frac{1}{3}\left( 1+2\frac{m_X^2}{m_h^2} 
             -\frac{2}{\cos \beta \sin \beta}
             \frac{m_{12}^2}{m_h^2}\right)
 \nonumber \\
 & \hspace*{100pt} \text{for } X=H,A,H^\pm,
 \nonumber\\
 c_{HXX} & = \frac{1}{3}\left( \tan \beta -\cot \beta \right)
  \nonumber \\
 & \hspace*{10pt} \cdot
              \left( \frac{m_H^2}{m_h^2}
             -\frac{1}{\cos \beta \sin \beta}
             \frac{m_{12}^2}{m_h^2}\right)
  \nonumber \\
 & \hspace*{10pt} \cdot
             \begin{cases}
               3 \text{\quad for } X=H\\
               1 \text{\quad for } X=A,H^\pm.
             \end{cases}
  \label{eq:tripleSM}
\end{align}

\section{Theoretical and Experimental Constraints}
\label{sec:constr}

For a detailed discussion of the theoretical and experimental constraints
included in our analysis we refer to our previous paper
\cite{Eberhardt:2013uba}. Here we only briefly list the included constraints and
comment on some updates and improvements which we made since the previous paper
\cite{Eberhardt:2013uba}. The theoretical constraints are
\begin{itemize}
\item positivity of the Higgs potential \cite{Deshpande:1977rw},
\item stability of the vacuum \cite{Barroso:2013awa} and
\item perturbativity of the Higgs self-couplings.
\end{itemize}

The perturbativity constraint is implemented in our analysis by requiring
\begin{equation}
  \|16\pi S\| < \Lambda_\text{max},
\end{equation}
where $S$ is the tree-level scattering matrix for Higgs and longitudinal gauge
bosons, as defined in \cite{Ginzburg:2005dt}, and the matrix norm $\|\cdot\|$ is
the magnitude of the largest eigenvalue. The unitarity of the tree-level
$S$-matrix, implemented in a partial-wave analysis of the two-particle Fock
states involving Higgs or longitudinal gauge bosons, requires
$\|S_{\phi\phi}\|\leq 1$, i.e.\ $\Lambda_\text{max}=16\pi$ \cite{Lee:1977eg}.
This bound has been studied for the 2HDM in \cite{Huffel:1980sk,
  Maalampi:1991fb, Kanemura:1993hm, Akeroyd:2000wc, Ginzburg:2005dt} and allows
the $\lambda_i$ in \eqref{eq:pot} to be as large as a few multiples of
$\pi$. For the SM, the unitarity analysis has been taken to the two-loop level
in \cite{Durand:1993vn}.  In \cite{Nierste:1995zx} upper bounds on the SM Higgs
coupling have been derived by imposing certain consistency conditions on the
perturbative series (order-by-order reduction of scheme and scale
dependences). Both analyses have resulted in significantly tighter upper bounds
on the on Higgs quartic coupling which, for the SM, corresponds roughly to
$\Lambda_\text{max}=2\pi$. In this paper we present results for two choices of
$\Lambda_\text{max}$: a \emph{loose} bound $\Lambda_\text{max}=16\pi$ and a
\emph{tight} bound $\Lambda_\text{max}=2\pi$. We advocate the use of the tight
bound. The loose bound is only included to show the sensitivity of certain
features in our plots to the implementation of the perturbativity constraint and
to facilitate comparisons with studies that only use tree-level unitarity to
constrain the size of the quartic Higgs couplings.

The experimental constraints included in our analysis are
\begin{itemize}
\item the signal strengths of the light $CP$ even Higgs boson $h$, as measured
  by the ATLAS \cite{ATLAS-CONF-2013-034} and CMS \cite{CMS-PAS-HIG-13-005}
  collaborations,
\item the CMS exclusion limits for a heavy Higgs decaying into $WW$, $ZZ$
  \cite{Chatrchyan:2013yoa} or $\tau\tau$ \cite{CMS-PAS-HIG-13-021} final
  states,
\item the full set of electroweak precision observables \cite{ALEPH:2005ab} (see
  \cite{Eberhardt:2013uba} for details) and
\item the flavour constraints relevant for the low $\tan\beta$ region, i.e.\ the
  constraints from the mass splitting $\Delta m_{B_s}$ of the neutral $B_s$
  system and the branching fraction $\br(B\to X_s\gamma)$.
\end{itemize}
Compared to our previous analysis \cite{Eberhardt:2013uba}, the signal strength
inputs have been improved by including the correlations between the different
Higgs production mechanisms. To this end, we compute for each final state
$X\in\{\gamma\gamma,ZZ,WW,b\bar b,\tau\tau\}$ the signal strengths
\begin{align}
    \mu_\text{ggF+ttH}^{X}
  &=\frac{(\sigma^\text{2HDM}_\text{ggF} + \sigma^\text{2HDM}_\text{ttH})
          \br(h\to X)^\text{2HDM}}
         {(\sigma^\text{SM}_\text{ggF} + \sigma^\text{SM}_\text{ttH})
           \br(h\to X)^\text{SM}},\nonumber\\
    \mu_\text{VBF+VH}^{X}
  &=\frac{(\sigma^\text{2HDM}_\text{VBF} + \sigma^\text{2HDM}_\text{VH})
          \br(h\to X)^\text{2HDM}}
         {(\sigma^\text{SM}_\text{VBF} + \sigma^\text{SM}_\text{VH})
           \br(h\to X)^\text{SM}},
\end{align}
where the subscripts `ggF', `ttH', `VBF' and `VH' stand for `gluon fusion',
`$t\bar t$ associated production', `vector boson fusion' and `Higgsstrahlung',
respectively and the superscripts indicate the model in which the corresponding
quantity is evaluated.  The central values, errors \emph{and correlations} of
the $\mu_\text{ggF+ttH}^{X}$ and $\mu_\text{VBF+VH}^{X}$ are then extracted from
Fig.~2 of \cite{ATLAS-CONF-2013-034} and Fig.~4 of \cite{CMS-PAS-HIG-13-005},
respectively. To compute the Higgs production cross sections and branching
fractions in the 2HDM we follow the recommendations of
\cite{Harlander:2013qxa}. In particular, we now compute the gluon fusion
production cross sections for 2HDM Higgs bosons by scaling the top-loop and
bottom-loop contributions \emph{and their interference} with appropriate 2HDM/SM
ratios of Higgs-top and Higgs-bottom couplings. The constraints from heavy Higgs
searches in the $WW$ and $ZZ$ final states were improved by using the combined
limits from Fig.~10 of \cite{Chatrchyan:2013yoa}. In addition, we now use the
model-independent limits from heavy Higgs searches in the $\tau\tau$ final state
from \cite{CMS-PAS-HIG-13-021} (median expected limits from Tabs.~7 and
8). Finally, our implementation of the electroweak precision observables now
uses the updated result \cite{Freitas:2012sy} for the SM prediction of
$R_b$.

Our analysis makes use of several public codes. For the computation of effective
$ggh$ and $h\gamma\gamma$ couplings and 2HDM contributions to electroweak
precision observables we use {\tt FeynArts 3.5}, {\tt FormCalc 7.0} and {\tt
  LoopTools 2.7} \cite{Hahn:1998yk, Hahn:2000kx, Hahn:2006qw}. The SM
contributions to the electroweak precision observables are computed with {\tt
  Zfitter 6.43} \cite{Bardin:1989tq, Bardin:1999yd, Arbuzov:2005ma} and
combined with the 2HDM contributions following the prescription in
\cite{Gonzalez:2011he}.  SM Higgs production cross sections and partial widths
were calculated with {\tt HIGLU 4.00} \cite{Spira:1995mt} and {\tt HDECAY 6.10}
\cite{Djouadi:1997yw, Butterworth:2010ym, Djouadi:2006bz}. The latter code
  was also used to calculate 2HDM branching fractions.  The fits were done with
the \textit{my}Fitter framework \cite{Wiebusch:2012en} which in turn uses {\tt
  Dvegas 2.0.3} \cite{Dvegas, Kauer:2001sp, Kauer:2002sn} for adaptive
parameter scans. The fit results were cross-checked with an independent
implementation in the CKMfitter framework \cite{Hocker:2001xe}.

\section{Fit Results}
\label{sec:results}

\subsection{Couplings to Vector Bosons and Fermions}
\label{sec:results:VVff}

\begin{figure}
  \centering
  \includegraphics{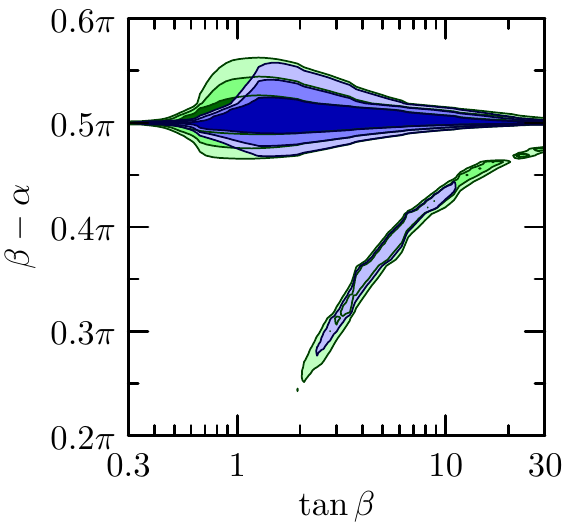}
  \caption{Allowed region in the $\tan\beta$-$(\beta-\alpha)$-plane. Shown are
    in blue the regions allowed at \unit{1}{\sigma} (dark), \unit{2}{\sigma}
    (medium) and \unit{3}{\sigma} (light) for the tight perturbativity bound
    ($\Lambda_\text{max}=2\pi$) as well as the corresponding regions for the
    loose perturbativity bound ($\Lambda_\text{max}=16\pi$) in green. Note that
    the lower strip is allowed at \unit{2}{\sigma} for
    $\Lambda_\text{max}=16\pi$ but only at \unit{3}{\sigma} for
    $\Lambda_\text{max}=2\pi$.}
  \label{fig:tb_bma}
\end{figure}

The tree-level couplings of the 2HDM Higgs bosons to vector bosons and fermions
are completely determined by the parameters $\tan\beta$ and $\beta-\alpha$.
Fig.~\ref{fig:tb_bma} shows the regions in the
$\tan\beta$-$(\beta-\alpha)$-plane which are allowed at \unit{1}{\sigma},
\unit{2}{\sigma} and \unit{3}{\sigma} for tight (blue) and loose (green)
perturbativity bound. For each combination of $\tan\beta$ and $\beta-\alpha$ we
perform a likelihood ratio test (LRT), i.e.\ we compute the difference $\Delta
\chi^2$ of minimum $\chi^2$-values obtained in the fit with all parameters free
and the fit with $\tan\beta$ and $\beta-\alpha$ fixed. The number of standard
deviations (sigmas) in this paper are two-sided $Z$-scores and are calculated in
the asymptotic limit, i.e.\ by assuming that $\Delta\chi^2$ has a
chi-square distribution with two degrees of freedom. Under this assumption the
$\Delta\chi^2$ values corresponding to \unit{1}{\sigma}, \unit{2}{\sigma} and
\unit{3}{\sigma} are 2.3, 6.2 and 11.8, respectively.

We see that at \unit{1}{\sigma} the value of $\beta-\alpha$ must be fairly close
to $\pi/2$---the alignment limit where the $h$ couplings are SM-like. At the
\unit{1}{\sigma} level deviations of more than $0.01\pi$ are only possible for
$\tan\beta$ between $0.8$ and $8$. The loose perturbativity bound admits
slightly larger deviations of $\beta-\alpha$ from $\pi/2$ for $\tan\beta<2$. For
$\tan\beta>2$ and $\beta-\alpha\in[0.45\pi,0.55\pi]$ there is essentially no
difference between the two perturbativity bounds. There is an additional
``island'' in the $\tan\beta$-$(\beta-\alpha)$-plane where $\beta-\alpha$ can be
as small as $0.3\pi$. For the loose perturbativity bound this island is allowed
at \unit{2}{\sigma}. For the tight bound it is only allowed at
\unit{3}{\sigma}. As discussed in \cite{Eberhardt:2013uba}, the best-fit points
in this island have an enhanced effective $ggh$ coupling and a reduced effective
$h\gamma\gamma$ coupling. They are allowed at \unit{1}{\sigma} by the Higgs
signal strengths but disfavoured by the flavour and electroweak precision
constraints, since they exhibit rather small values of $m_{H^\pm}$.

\subsection{Triple Higgs Couplings}
\label{sec:results:hhh}

\begin{figure*}
  \centering
  \includegraphics{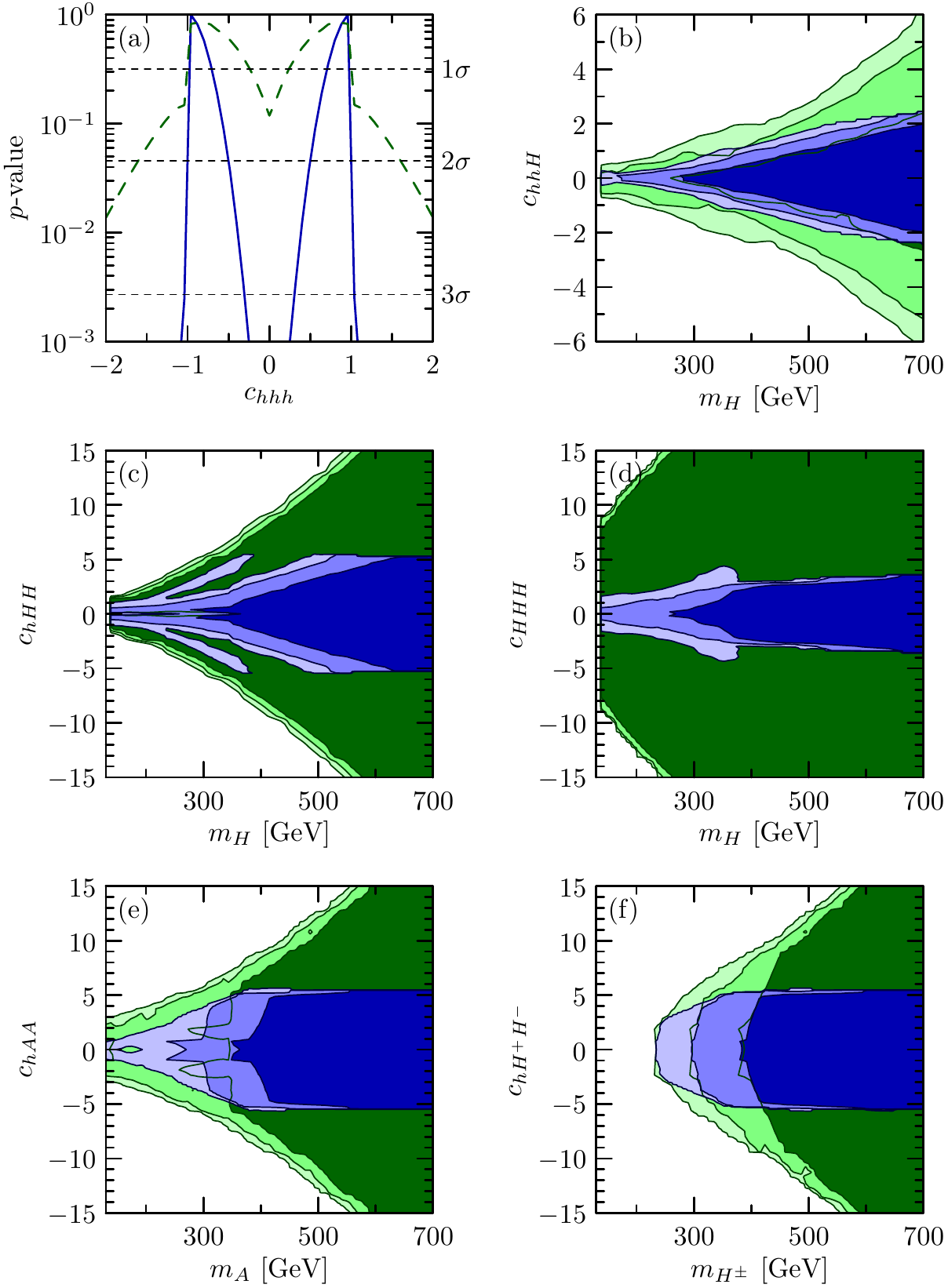}
  \caption{(a) $p$-value of the 2HDM with a fixed 2HDM/SM ratio $c_{hhh}$ of
    triple $h$ couplings (see \eqref{eq:triple} for the complete expression) as
    a function of $c_{hhh}$. The solid line corresponds to the tight
    perturbativity bound ($\Lambda_\text{max}=2\pi$) and the dashed line to the
    loose bound ($\Lambda_\text{max}=16\pi$). (b)-(f) Allowed ranges for the
    2HDM/SM triple Higgs coupling ratios as a function of the corresponding
    heavy Higgs mass. Shown are in blue the regions allowed at \unit{1}{\sigma}
    (dark), \unit{2}{\sigma} (medium) and \unit{3}{\sigma} (light) for the tight
    perturbativity bound ($\Lambda_\text{max}=2\pi$) as well as the
    corresponding regions for the loose perturbativity bound
    ($\Lambda_\text{max}=16\pi$) in green. The $p$-values in (a) were computed
    by assuming a chi-square distribution for the test statistic with one degree
    of freedom. For the significances in (b)-(f) we assumed a chi-square
    distribution with two degrees of freedom.}
  \label{fig:cXXX}
\end{figure*}

>From \eqref{eq:tripleSM} we know that $c_{hhh}=1$ in the alignment limit.  Thus,
deviations of the triple-$h$ coupling from its SM value, i.e.\ $c_{hhh}\neq 1$,
imply $\beta-\alpha\neq \pi/2$ and therefore only appear in conjunction with
corresponding deviations of $h$ couplings to fermions and gauge bosons.  The
observed Higgs signal strengths can therefore be used to constrain the possible
deviations of the triple $h$ coupling in the 2HDM. To calculate these
constraints, we perform a LRT of the 2HDM with $c_{hhh}$ fixed.
Fig.~\ref{fig:cXXX}a shows the $p$-values of the test as a function of the value
of $c_{hhh}$. The $p$-values were computed by assuming a chi-square distribution
with one degree of freedom for $\Delta\chi^2$. For the loose perturbativity
bound (dashed line) we see that the magnitude of $c_{hhh}$ can reach 1.7 at
\unit{2}{\sigma}. A reduction of the coupling to zero is also possible at
\unit{2}{\sigma}. At \unit{1}{\sigma} (and for $\Lambda_\text{max}=16\pi$) the
magnitude of $c_{hhh}$ cannot exceed 1 and must be bigger than $0.25$.
The best-fit scenarios outside the \unit{1}{\sigma} region for the loose
perturbativity bound feature quartic couplings
$\lambda_i$ (c.f.~\eqref{eq:pot}) with magnitude above $10$. If we impose the
tight perturbativity bound the magnitude of $c_{hhh}$ can no longer exceed $1$
at \unit{3}{\sigma}.  It can be reduced to $0.72$ at \unit{1}{\sigma}, to
$0.56$ at \unit{2}{\sigma} and to $0.4$ at \unit{3}{\sigma}.

The coupling ratio $c_{hhH}$ vanishes in the alignment limit and is thus also
strongly constrained by the Higgs signal strengths. Fig.~\ref{fig:cXXX}b shows
the ranges for $c_{hhH}$ allowed at the $1$, $2$ and \unit{3}{\sigma} levels as
a function of $m_H$. For $m_H\gtrsim\unit{400}{GeV}$ the \unit{1}{\sigma}
contour for the loose perturbativity bound coincides with the corresponding
contour for the tight perturbativity bound. The best-fit points inside this
contour are close to the alignment limit and mainly constrained by the Higgs
signal strengths. For the tight perturbativity bound the allowed region does not
increase significantly when we go to the $2$ or \unit{3}{\sigma} level, but it
does for the loose perturbativity bound. This is due to the fact that the loose
perturbativity bound allows scenarios far away from the alignment limit at
\unit{2}{\sigma}, as seen in Fig.~\ref{fig:tb_bma}.

The remaining triple Higgs couplings are not fixed in the alignment limit and
their allowed range is mainly determined by the perturbativity
requirement. Consequently, the allowed ranges for the tight and loose
perturbativity bounds differ substantially. Figs.~\ref{fig:cXXX}c-f show the
ranges allowed at $1$, $2$ and \unit{3}{\sigma} as a function of the heavy Higgs
mass. While the loose perturbativity bound allows enhancement factors of $20$ or
more in the displayed mass range, the enhancement factors for the tight
perturbativity bound cannot exceed $6$ in magnitude. The symmetry of the plots
is due to the following property of the triple Higgs couplings: if we shift
$\beta-\alpha$ by $\pi$ and keep all other parameters in \eqref{eq:params} fixed
all triple Higgs couplings flip their sign but keep the same magnitude.
Such a shift in $\beta-\alpha$ corresponds to a field re-definition
$(h,H)\to(-h,-H)$ and therefore has no physical consequence.

\subsection{Branching Fractions}
\label{sec:results:br}

\begin{figure*}
  \centering
  \includegraphics{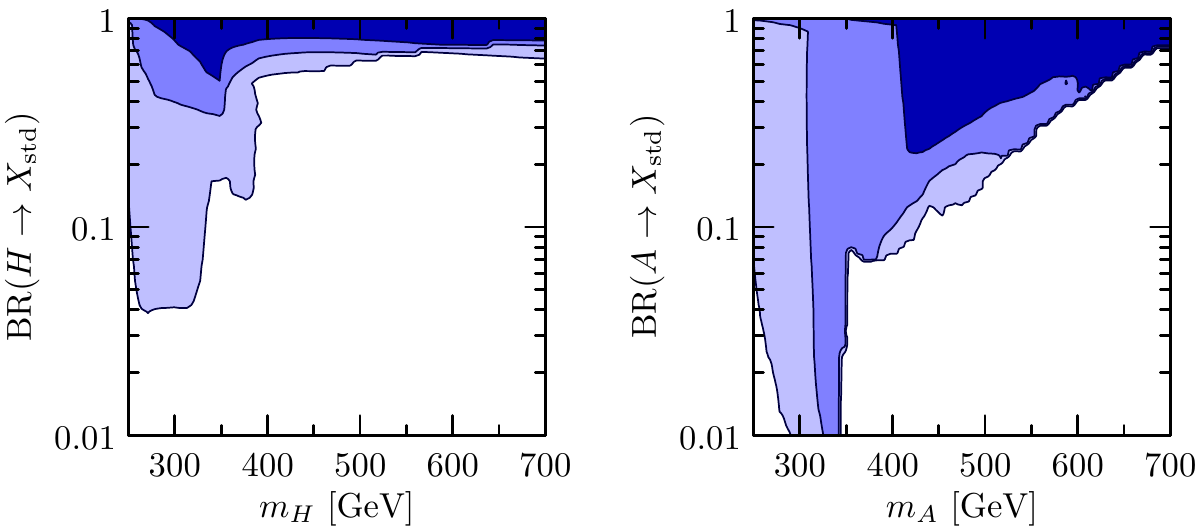}
  \caption{Allowed ranges for the branching fractions of heavy neutral Higgs
    bosons into `standard' final states (i.e.\ states which do not contain
    another Higgs boson) as a function of the corresponding heavy Higgs
    mass. Shown are the regions allowed at \unit{1}{\sigma} (dark),
    \unit{2}{\sigma} (medium) and \unit{3}{\sigma} (light) for the tight
    perturbativity bound ($\Lambda_\text{max}=2\pi$).}
  \label{fig:brVVff}
\end{figure*}

The search for additional Higgs resonances in the 2HDM is complicated by the
possible existence of tree-level decays into lighter Higgs bosons like $H\to hh$
or $A\to Zh$. If kinematically allowed, these decays compete with the standard
decay modes $H,A\to X_\text{std}$, where $X_\text{std}$ stands for all
`standard' final states which do not contain another Higgs boson (i.e.\ $gg$,
$\gamma\gamma$, $WW$, $ZZ$, $Z\gamma$, $t\bar t$, $b\bar b$, $\tau\tau$
etc.). Of course the non-standard decays may also be regarded as alternative
discovery modes for the heavy Higgs bosons or as additional sources of
background for the observation of the light SM-like Higgs. To estimate the
possible size of the non-standard branching fractions of the heavy Higgs bosons
we show the $1$, $2$ and \unit{3}{\sigma} allowed regions in the $m_H$-$\br(H\to
X_\text{std})$ plane in Fig.~\ref{fig:brVVff}a. Only the results for
the tight perturbativity bound are displayed.

To achieve the largest non-standard branching fractions the mass of the decaying
Higgs boson has to be below the $t\bar t$ threshold, so that the non-standard
decays do not have to compete with the large $t\bar t$ partial width. In this
mass region the branching fraction $\br(H\to X_\text{std})$ can drop below
$40\%$ at the \unit{2}{\sigma} level. This is due to the non-standard $H\to hh$
decay. The branching fraction $\br(A\to X_\text{std})$ can drop below $1\%$ at
the \unit{2}{\sigma} level if $m_A$ is in the narrow window between
$\unit{320}{GeV}$ and $2m_t$. This is due to a non-trivial interplay of all
constraints in our fit: since couplings involving the light Higgs cannot be
enhanced very much, the largest non-standard partial widths are obtained for
heavy-to-heavy decay modes like $A\to ZH$. For this decay to be open we need a
mass splitting of more than \unit{90}{GeV} between $A$ and $H$.  To close the
competing $A\to t\bar t$ decay channel, $m_A$ must be below
$2m_t\approx\unit{350}{GeV}$, which puts $m_H$ below \unit{260}{GeV}.  But
electroweak precision constraints require one of the two heavy neutral Higgs
bosons to be almost degenerate with the charged Higgs, and flavour constraints
force the charged Higgs mass to be above approximately \unit{320}{GeV}. Since
the charged Higgs cannot be degenerate with $H$ (as $m_H$ is below
\unit{260}{GeV}) it must be degenerate with $A$, and this means $m_A$ must be
between \unit{320}{GeV} and $2m_t$ to achieve the largest $A\to ZH$ partial
width. For $m_A$ above $2m_t$ the branching fraction $\br(A\to X_\text{std})$
can still be below $8\%$ at the \unit{2}{\sigma} level, but the lower limit
increases with $m_A$. For $m_A\gtrsim\unit{400}{GeV}$ the $A\to H^\pm W^\mp$
decays are possible without violating the constraint on $m_{H^\pm}$ from
$\br(\bar B\to X_s\gamma)$.  As a result, standard branching fractions as low as
$30\%$ are allowed at \unit{1}{\sigma} for $m_A\approx\unit{400}{GeV}$.

\section{Phenomenology}
\label{sec:phen}

\subsection{Benchmarks}
\label{sec:phen:bench}

The previous section has shown that today's measurements place severe
constraints on parameters of the 2HDM of type II to be probed in Run II of the
LHC and at a future linear collider. Our results can serve as input for
studies of collider signals in the type-II 2HDM. In practice, these are
only feasible for a small number of benchmark points. The benchmarks
should maximise the collider signal under investigation while being in agreement
with all existing constraints. Therefore, the criteria by which the benchmark
points should be selected depend on the physics one wants to investigate. In
searches for (non-resonant) Higgs pair production one is generally interested in
scenarios with large triple Higgs couplings, while searches for heavy Higgs
resonances with non-standard decay modes need benchmarks where the branching
fractions of the non-standard decays are as large as possible. In this section
we provide benchmark points for the type-II 2HDM which are tailored for these
two types of new physics searches.

Our benchmarks for large triple-Higgs couplings are summarised in
Tab.~\ref{tab:cbench}. The benchmarks were constructed from best-fit parameters
corresponding to points inside the \unit{2}{\sigma} allowed regions in
Figs.~\ref{fig:cXXX}a-e and as close as possible to their edges. As such, they
agree with all constraints in our fit at the \unit{2}{\sigma} level. Benchmark
a-1 corresponds to Fig.~\ref{fig:cXXX}a and is chosen to minimise the absolute
value of $c_{hhh}$.  Benchmarks b-1 to b-4 correspond to Fig.~\ref{fig:cXXX}b
and maximise $|c_{hhH}|$ for different fixed values of $m_H$. The remaining
benchmarks in Tab.~\ref{tab:cbench} were constructed from
Figs.~\ref{fig:cXXX}c-e in an analogous way.

\begin{table*}
  \centering
  \begin{tabular}{lrrrrrr}
    \hline\hline
      & $\tan\beta$ & $(\beta-\alpha)/\pi$ & $m_H\ \text{[GeV]}$ & $m_A\ \text{[GeV]}$
      & $m_{H^\pm}\ \text{[GeV]}$ & $m_{12}^2\ \text{[GeV${}^2$]}$ \\
    \hline
    a-1  & 1.50 &    0.529 & 700 & 700 &  670 & 180000 \\
    \hline
    b-1  & 2.52 &    0.511 & 200 & 383 &  383 &  14300 \\
    b-2  & 2.23 &    0.525 & 300 & 444 &  445 &  32700 \\
    b-3  & 1.73 &    0.533 & 400 & 502 &  503 &  62900 \\
    b-4  & 1.74 &    0.533 & 600 & 619 &  579 & 126000 \\
    \hline
    c-1  & 2.22 &    0.505 & 200 & 337 &  337 &  12000 \\
    c-2  & 1.88 &    0.509 & 300 & 361 &  365 &  27800 \\
    c-3  & 1.49 &    0.518 & 400 & 350 &  407 &  47100 \\
    c-4  & 1.25 &    0.522 & 600 & 491 &  600 & 123000 \\
    \hline
    d-1  & 2.78 &    0.503 & 200 & 319 &  320 &  10400 \\
    d-2  & 2.17 &    0.507 & 300 & 350 &  347 &  26500 \\
    d-3  & 1.85 &    0.503 & 400 & 350 &  404 &  53100 \\
    d-4  & 2.40 &    0.520 & 600 & 634 &  587 & 114000 \\
    \hline
    e-2  & 5.34 &    0.502 & 250 & 300 &  307 &  10700 \\
    e-3  & 4.90 &    0.502 & 229 & 400 &  399 &  10300 \\
    e-4  & 6.45 &    0.502 & 498 & 600 &  601 &  37530 \\
    \hline\hline
  \end{tabular}
  \caption{Benchmark scenarios with enhanced/reduced triple Higgs
    couplings. Benchmark a-1 approximates the best-fit scenario with reduced
    $hhh$ coupling at the edge of the \unit{2}{\sigma} interval in
    Fig.~\ref{fig:cXXX}a. Benchmarks b-1 to b-4 approximate the best-fit
    scenarios associated with points on the \unit{2}{\sigma} contour in
    Fig.~\ref{fig:cXXX}b.  Benchmarks c-1 to e-4 are related to
    Figs.~\ref{fig:cXXX}c, \ref{fig:cXXX}d and \ref{fig:cXXX}e in an analogous
    way. All points are allowed at the \unit{2}{\sigma} level.}
  \label{tab:cbench}
\end{table*}

Tab.~\ref{tab:brbench} contains benchmarks featuring large non-standard decay
rates of the heavy Higgs bosons $H$ and $A$. They were constructed from best-fit
points inside the \unit{2}{\sigma} contours in Fig.~\ref{fig:brVVff} and as
close as possible to their edges. The branching fractions of $H$ and $A$ for
these benchmark points are given in Tabs.~\ref{tab:brH} and \ref{tab:brA},
respectively. We see that for scenarios that minimize $\br(H\to X_\text{std})$
(at the \unit{2}{\sigma} level) the largest non-standard decay rate is $\br(H\to
hh)$ while for scenarios that minimize $\br(A\to X_\text{std})$ the largest
non-standard decay rate is $\br(A\to ZH)$.
\begin{table*}
  \centering
  \begin{tabular}{lrrrrrr}
    \hline\hline
      & $\tan\beta$ & $(\beta-\alpha)/\pi$ & $m_H\ \text{[GeV]}$ & $m_A\ \text{[GeV]}$
      & $m_{H^\pm}\ \text{[GeV]}$ & $m_{12}^2\ \text{[GeV${}^2$]}$ \\
    \hline
    H-1  & 1.75 &    0.522 & 300 & 441 &  442 &  38300 \\
    H-2  & 2.00 &    0.525 & 340 & 470 &  471 &  44400 \\
    H-3  & 4.26 &    0.519 & 450 & 546 &  548 &  43200 \\
    H-4  & 4.28 &    0.513 & 600 & 658 &  591 &  76900 \\
    \hline
    A-1  & 4.61 &    0.505 & 346 & 300 &  345 &  23600 \\
    A-2  & 2.74 &    0.503 & 131 & 340 &  339 &   6200 \\
    A-3  & 7.02 &    0.508 & 290 & 450 &  446 &  11700 \\
    A-4  & 7.44 &    0.504 & 490 & 600 &  598 &  31620 \\
    \hline\hline
  \end{tabular}
  \caption{Benchmark scenarios with large non-standard decay rates of heavy
    Higgs bosons. Benchmarks H-1 to H-4 approximate the best-fit scenarios
    associated with points on the lower \unit{2}{\sigma} contour in
    Fig.~\ref{fig:brVVff}a. Benchmarks A-1 to A-4 are related to
    Fig.~\ref{fig:brVVff}b in an analogous way. All points have a $\Delta\chi^2$
    less than $6.2$, i.e.\ are allowed at \unit{2}{\sigma} in a LRT with two
    degrees of freedom.}
  \label{tab:brbench}
\end{table*}
\begin{table}
  \centering
  \begin{tabular}{lcrrrrrrrr}
    \hline\hline
         & $m_H\ \text{[GeV]}$ & $t\bar t$ & $b\bar b$ & $\tau\tau$ &
         $WW$ & $ZZ$ & $gg$ & $hh$ \\
    \hline
    H-1  & 300 &  0.1 & 10.6 &  1.2 & 18.9 &  8.4 &  1.8 & 59.0 \\
    H-2  & 340 &  1.4 &  7.0 &  0.8 & 17.8 &  8.1 &  1.2 & 63.7 \\
    H-3  & 450 & 51.6 &  9.2 &  1.1 &  6.0 &  2.8 &  0.2 & 29.0 \\
    H-4  & 600 & 64.0 &  7.3 &  0.9 &  4.4 &  2.1 &  0.2 & 21.1 \\
    \hline
    A-1  & 346 &  1.5 & 81.0 &  9.5 &  1.5 &  0.7 &  0.6 &  5.1 \\
    A-2  & 131 &      & 90.4 &  8.8 &      &      &  0.6 &      \\
    A-3  & 290 &      & 85.2 &  9.7 &  1.0 &  0.5 &  0.2 &  3.4 \\
    A-4  & 490 & 32.5 & 56.3 &  7.0 &  0.6 &  0.3 &  0.1 &  3.0 \\
    \hline\hline
  \end{tabular}
  \caption{Branching fractions (in percent) of the heavy scalar Higgs
    boson $H$ for the benchmarks from Tab.~\ref{tab:brbench}}
  \label{tab:brH}
\end{table}

\begin{table}
  \centering
  \begin{tabular}{lcrrrrrr}
    \hline\hline
         & $m_A\ \text{[GeV]}$ & $t\bar t$ & $b\bar b$ & $\tau\tau$ & $gg$
         & $Zh$ & $ZH$ \\
    \hline
    H-1  & 441 & 74.9 &  0.3 &      &   0.3 &  1.1 & 23.4 \\
    H-2  & 470 & 77.4 &  0.5 &  0.1 &   0.3 &  2.2 & 19.6 \\
    H-3  & 546 & 78.4 &  9.1 &  1.2 &   0.3 &  8.2 &  3.0 \\
    H-4  & 658 & 82.8 &  9.2 &  1.2 &   0.2 &  6.3 &      \\
    \hline
    A-1  & 300 &  0.1 & 88.7 & 10.1 &   0.4 &  0.5 &      \\
    A-2  & 340 &  0.4 &  0.8 &  0.1 &   0.1 &      & 98.6 \\
    A-3  & 450 &  9.6 &  9.2 &  1.1 &       &  0.3 & 79.7 \\
    A-4  & 600 & 32.5 & 33.8 &  4.4 &   0.1 &  0.6 & 28.7 \\
    \hline\hline
  \end{tabular}
  \caption{Branching fractions (in percent) of the pseudoscalar Higgs boson $A$
    for the benchmarks from Tab.~\ref{tab:brbench}}
  \label{tab:brA}
\end{table}

\subsection{Higgs Pair Production in the Gluon Fusion Channel}
\label{sec:phen:hpair}

As we have seen in Section 4 the triple Higgs coupling between the light Higgs
state cannot be enhanced. However, this does not mean that the gluon fusion
cross section $\sigma(gg\to hh)$ cannot be enhanced with respect to the SM. In
the second diagram of Fig.~\ref{fig:diag}, not only the light Higgs state $h$
contributes in the propagator, but also the heavy Higgs state $H$. For $m_H >
2m_h$ this diagram represents a resonant $H$ production with a subsequent
non-standard decay into $hh$, while for $m_H < 2m_h$ it gives a non-resonant
contribution to the $hh$ production cross section.

In the non-resonant case we expect large deviations of $\sigma(gg\to hh)$ from
its SM value when $|c_{hhH}|$ is large and $m_H$ just below the $hh$
threshold. However, for $m_H<2m_h$ the absolute value of $c_{hhH}$ is
constrained to be below approximately $0.17$ at the \unit{2}{\sigma} level, as
can be seen in Fig.~\ref{fig:cXXX}b. To study such a scenario in detail we
computed $\sigma(gg\to hh)$ for benchmark b-1 (see Tab.~\ref{tab:cbench}).
Using the set-up described in Ref. \cite{Baglio:2012np} and a modified version
of \texttt{HPAIR} \cite{Dawson:1998py} with the type-II 2HDM implemented, we
obtain $\sigma(gg\to hh)_\text{2HDM,b-1}=\unit{30.3}{fb}$ at the LHC with
$\unit{14}{TeV}$.  This is a reduction of $10\%$ compared to the SM value
$\sigma(gg\to hh)_\text{SM}=\unit{33.5}{fb}$, and is due to the fact that the
triangle diagram (c.f. Fig.~\ref{fig:diag}) involving the heavy Higgs $H$
interferes destructively with the box diagram and the triangle diagram involving
the light Higgs $h$. In addition, the $hhh$ and $ht\bar t$ couplings are
slightly reduced in this scenario. These results indicate that a large
enhancement is not possible in the non-resonant case.

\begin{figure}
  \centering
  \includegraphics[scale=0.8]{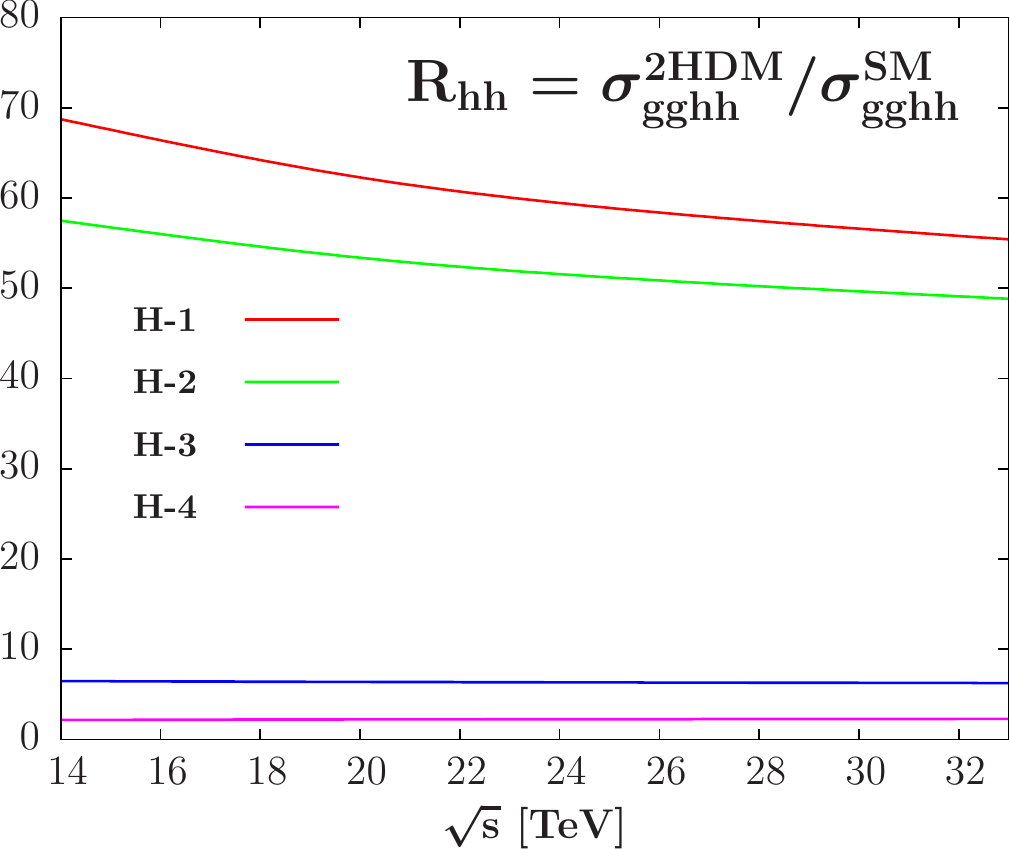}
  \caption{Ratio between the gluon fusion cross section $\sigma(gg\to
    hh)$ at the LHC in 2HDM of type II using benchmark scenarios of
    type H in Tab.~\ref{tab:brbench}, and the same cross section
    but within the SM, as a function of the center-of-mass energy
    $\sqrt{s}$ (in TeV).}
  \label{fig:ggHH}
\end{figure}

In the resonant case we expect large enchancements of $\sigma(gg\to hh)$ when
the $H\to hh$ decay has a large branching fraction. This situation is realised
in benchmarks H-1 to H-4 (see Tab.~\ref{tab:brbench}).  Fig.~\ref{fig:ggHH}
shows the 2HDM/SM ratio of the $gg\to hh$ cross section as a function of the
hadronic centre of mass energy for these benchmarks. The magnitude of $c_{hhh}$
is close to $1$ for all these benchmarks, which means that the contribution from
the diagram with the $s$-channel $h$ exchange is SM-like. We see that the $gg\to
hh$ cross section can be enhanced by more than a factor 50 for benchmarks H-1
and H-2 and that benchmark H-3 still gives a factor 6
enhancement. 

We also investigated the possibility that the only hints for heavy Higgs
resonances at the \unit{14}{TeV} run of the LHC after \unit{300}{\power{fb}{-1}}
of data are in the $hh$ final state.  Using an extrapolation of the current
limits for heavy Higgs resonances in the $WW$, $ZZ$ and $\tau\tau$ channels
based on the ratio of expected signal events we find that scenarios with heavy
Higgs masses below the $t\bar t$ threshold can be ruled out (or discovered) by
the standard searches. The ILC in its first stage at $\sqrt{s}=\unit{250}{GeV}$
would impose even stronger bounds on additional Higgs resonances below the
$t\bar t$ threshold, but provide little information about heavier
resonances. Benchmarks H-3 and H-4 can easily evade the expected limits from
both ILC and the LHC at 14 TeV as they have all heavy Higgs bosons above the
$t\bar t$ threshold. However, as seen in Fig.~\ref{fig:ggHH} they still lead to
a noticable enhancement in the Higgs pair production rate.

\section{Conclusions}
\label{sec:conclusions}

The existing data from flavour, electroweak precision and Higgs physics imposes
severe constraints on extended Higgs sectors and can be used to to put limits on
the possible size of signals in future collider searches. In this paper we
studied the bounds on triple Higgs couplings and branching fractions of
non-standard decays of heavy Higgs bosons (i.e. decays into other Higgs bosons)
in the context of the $CP$ conserving two-Higgs-doublet-model of type II. Our
analysis uses current experimental constraints from flavour, electroweak
precision and Higgs data as well as theoretical constraints such as positivity
of the Higgs potential, vacuum stability and perturbativity. While our results
were presented for two different implementations of the perturbativity bound, we
advocate the use of the ``tight'' bound which requires the magnitude of the
eigenvalues of the tree-level $\phi\phi\to\phi\phi$ $S$-matrix to be below
$1/8$.

With regard to the questions (i) to (iv) raised in the introduction, our
results may be summarised as follows:
\begin{itemize}
\item[i)] assuming that the observed Higgs resonance is the light $CP$-even Higgs
  boson $h$ of the 2HDM, the data on Higgs signal strengths pushes the
  parameters of the 2HDM close to the alignment limit, where the couplings of
  $h$ are SM-like. In this limit, also the $hhh$ coupling is SM-like and the
  $hhH$ coupling vanishes. Deviations of these couplings from the alignment
  limit are therefore strongly constrained by the observed Higgs signal
  strengths.  Using all available theoretical and experimental constraints we
  have shown that an enhancement of the $hhh$ coupling is forbidden at the
  \unit{3}{\sigma} level, while a reduction by a factor of $0.56$ is possible at
  the \unit{2}{\sigma} level (Fig.~\ref{fig:cXXX}a).
\item[ii)] The allowed range for the $hhH$ coupling increases linearly with
  $m_H$ and then saturates at $2.5$ times the size of the SM $hhh$ coupling
  (Fig.~\ref{fig:cXXX}b).  The $hHH$, $HHH$, $hAA$ and $hH^+H^-$ couplings are
  not fixed in the alignment limit and are therefore mostly constrained by the
  requirement of perturbativity. Their allowed ranges also increase and then
  saturate with the mass of the corresponding heavy Higgs boson. Using the tight
  perturbativity bound we find that these couplings can be at most $5.5$ times
  as large as the SM $hhh$ coupling (Figs.~\ref{fig:cXXX}c-f).

  The results of our fits were condensed into a set of benchmark points
  (Tab.~\ref{tab:cbench}) which agree with present experimental data at the
  \unit{2}{\sigma} level and feature large triple Higgs couplings. These
  benchmarks can be used to study collider signatures for processes like Higgs
  pair production, where large triple Higgs couplings lead to enhanced signals.
\item[iii)] To estimate the possible suppression of heavy Higgs signals due to
  non-standard decay modes (i.e.\ decays involving another Higgs boson) we have
  determined the allowed range of the standard branching fractions $\br(H\to
  X_\text{std})$ and $\br(A\to X_\text{std})$ (with $X_\text{std}=gg, WW, ZZ,
  t\bar t, b\bar b, \tau\tau, \ldots$) as a function of the decaying Higgs
  mass. We find that the strongest suppression of the standard decay modes
  occurs for heavy Higgs masses below the $t\bar t$ threshold. If $m_H$ is below
  $2m_t$ the branching fraction $\br(H\to X_\text{std})$ can be reduced to
  $40\%$ at the \unit{2}{\sigma} level due to the competing $H\to hh$ decay.
  Above the $t\bar t$ threshold values of $\br(H\to X_\text{std})$ as low as
  $70\%$ are still possible (Fig.~\ref{fig:brVVff}).  The branching fraction
  $\br(A\to X_\text{std})$ for the pseudo-scalar $A$ can be much smaller when
  competing heavy-to-heavy decays like $A\to ZH$ or $A\to W^\pm H^\mp$ are
  kinematically allowed.  This is due to the fact that the relevant triple Higgs
  couplings are not as strongly constrained as the couplings involving the light
  Higgs $h$.  The strongest reductions of $\br(A\to X_\text{std})$ which are
  allowed at the \unit{2}{\sigma} level occur when the $A\to ZH$ decay is open.
  for $m_A$ between \unit{320}{GeV} and $2m_t$ this can lead to a reduction
  below $1\%$.

  To facilitate further studies of 2HDM scenarios with large non-standard
  branching fractions we have provided benchmark points with maximal
  non-standard decay rates (Tab.~\ref{tab:brbench}).
\item[iv)] To study the effect of the heavy Higgs boson $H$ in light Higgs pair
  production we have calculated the $gg\to hh$ cross section at the
  \unit{14}{TeV} LHC for our benchmark points b-1 and H-1 to H-4. For benchmark
  b-1 the heavy Higgs only gives a non-resonant contribution and actually leads
  to a reduction of the cross section compared to the SM. This is due to a
  destructive interference between the triangle diagram
  (c.f.\ Fig.~\ref{fig:diag}) and the diagrams involving only the light Higgs
  $h$. For benchmarks H-1 to H-4 the heavy Higgs is produced on-shell and then
  decays into an $hh$ pair. In scenarios H-1 and H-2 the SM $gg\to hh$ cross
  section is enhanced by more than a factor 50. For H-3 we still find an
  enhancement by a factor 6 (Fig.~\ref{fig:ggHH}). 

  Using a naive extrapolation of the current limits for heavy Higgs resonances
  we find that benchmarks H-1 and H-2 can be verified or ruled out by standard
  $A\to\tau\tau$ searches with \unit{300}{\power{fb}{-1}} of data at a
  \unit{14}{TeV} LHC. However, for benchmarks H-3 and H-4 the heavy Higgs bosons
  would still be invisible to these searches and the only hint for their
  existence might be the enhanced $gg\to hh$ cross section.
\end{itemize}

It is often stated that a measurement of the $hhh$ coupling may reveal new
physics. In this paper we have found that this statement does not apply for the
discrimination of the SM and the 2HDM of type II.  Current data on the $h$
couplings to vector bosons and fermions (together with flavour and electroweak
precision data) already constrain the maximally possible deviation of the 2HDM
$hhh$ coupling from the SM limit to be of the order of the experimental accuracy
expected from \unit{300}{\power{fb}{-1}} of \unit{14}{TeV} LHC data. This means
that substantial improvements in the SM theoretical predictions and in
experimental tools are required to discriminate between the type-II 2HDM and the
SM by means of the triple $h$ coupling. While the $h$ pair production cross
section in gluon fusion may be significantly enhanced in the 2HDM (see (iv)
above), this feature is mostly independent of the size of the $hhh$ coupling.

\begin{acknowledgments}
We thank Ansgar Denner, Rui Santos and C\'edric Weiland for fruitful
discussions and the CKMfitter group for allowing us to use their
statistical analysis framework. The research leading to these results
has received funding from the European Research Council under the
European Union's Seventh Framework Programme (FP/2007-2013) / ERC
Grant Agreement n.~279972.  We also acknowledge support by DFG through
grants NI1105/2-1 and LE1246/9-1.  J.B. acknowledges support by DFG
through the SFB/TR-9 Computational Particle Physics.
\end{acknowledgments}

\subsection*{Note added}
After the completion of our work a similar study \cite{Dumont:2014wha} appeared which focusses on the prospects of direct observation of the heavy Higgs bosons at the \unit{14}{TeV} LHC run in the context of 2HDM of type I and II and also discusses the case that the observed Higgs resonance is the heavy $CP$-even Higgs $H$. Our results concerning the allowed range of the $hhh$ coupling in the 2HDM of type II have been confirmed by the results of this new paper.

\bibliographystyle{apsrev4-1}
\bibliography{triplehiggs}

\end{document}